# 3D MIMO Scheme for Broadcasting Future Digital TV in Single Frequency Networks

Youssef Nasser, Jean-François Hélard and Matthieu Crussière

**Abstract**

This letter introduces a 3D space-time-space block code for future terrestrial digital TV systems. The code is based on a double layer structure designed for inter-cell and intra-cell transmissions in single frequency networks. Without increasing the complexity of the receiver, the proposed code is very efficient to cope with equal and unequal received powers in single frequency network scenarios.

## 1    Introduction

Nowadays, one of the most promising technologies for the second generation of future terrestrial digital TV, concerned with flexibility, high bit rate and, portable and mobile reception is the combination of multiple-input multiple-output (MIMO) and orthogonal frequency division multiplexing (OFDM) techniques. To increase area coverage, single frequency networks (SFN) [1] are used for broadcasting terrestrial digital TV. SFN are based on the simple addition of lower power transmitters at various sites throughout the coverage area. In an SFN, several transmitters transmit at the same moment the same signal on the same frequency. Because it is desirable to deploy SFN with lower transmitted powers, increased bit rates and better performance, new MIMO-OFDM systems have to be designed to ensure such transmission conditions.

In this letter, we present a 3-dimension (3D) space-time-space block code (STSBC) for MIMO-OFDM systems in SFN with mobile and portable reception. The use of a second space dimension is due to SFN. The proposed code is based on the combination of 2 layers: one layer corresponds to an inter-cell ST coding, the second corresponds to an intra-cell ST coding. In the following, we first present the scenario of mobile and portable reception with single layer reception. Then, we introduce our proposed code as a double layer code and adapt it to the SFN environment.

## 2    3D code

Consider a MIMO-OFDM communication system using ($2 \times M_T$) transmit antennas (Tx) and $M_R$ receive antennas (Rx) for a downlink communication. In this letter, we propose to apply a distributed MIMO scheme in an SFN architecture. Such a system could be implemented on 2 different sites using $M_T$ Tx by site as shown in Figure 1. The transmission could therefore be seen as a double layer scheme in the space domain. The first layer is seen between the 2 sites separated by $D$ km (distributed MIMO scheme). The second layer is seen between the antennas separated by $d$ m within one site. For the first layer, an STBC encoding scheme is applied between the 2 signals transmitted by each site antenna. In the second layer, we use a second STBC encoder for each subset of $M_T$ signals transmitted from the same site. For the first layer, the STBC encoder takes $L$ sets of $Q$ data complex symbols each $(s_1,…,s_Q)$ and transforms them into a $2 \times U$ output matrix according to the STBC scheme. In the second layer (the second step), the encoder transforms each component of the first layer matrix into $M_T \times T$ output matrix according to the second layer STBC scheme. The number of rows of the encoding matrix in the first layer is equal to two since, the STBC scheme is applied between the signals of two different sites. The output signal of each site is fed to $M_T$ OFDM modulators, each using $N$ sub-carriers. The reader could construct a double layer Alamouti code, for example, by considering 2 sets of 2 symbols each and then, by applying Alamouti encoding between the 2 symbols' sets and another Alamouti encoding between the signals in each site. More generally, the double layer encoding matrix is described by:

$$\mathbf{X}^{(1)} = \begin{pmatrix} \mathbf{X}^{(2)}_{11} & \cdots & \mathbf{X}^{(2)}_{1U} \\ \mathbf{X}^{(2)}_{21} & \cdots & \mathbf{X}^{(2)}_{2U} \end{pmatrix}$$

$$\mathbf{X}^{(2)}_{ij} = \begin{pmatrix} f_{ij,11}(s_1,…s_Q) & \cdots & f_{ij,1T}(s_1,…s_Q) \\ \vdots & \ddots & \vdots \\ f_{ij,M_T 1}(s_1,…s_Q) & \cdots & f_{ij,M_T T}(s_1,…s_Q) \end{pmatrix} \quad (1)$$

In (1), the superscript indicates the layer, $f_{ij,mt}(s_1,…s_Q)$ is a function of the input complex symbols $s_q$ and depends on the STBC encoder scheme. The time dimension of the resulting 3D code is equal to $U \times T$ and the resulting coding rate is $R = \frac{Q \times L}{U \times T}$. In order to have a fair analysis and comparison between different STBC codes, the signal power at the output of the ST encoder at each site is normalized by $2 \times M_T$. In the following, we will compare different STBC schemes assuming that a portable or mobile terminal receives signals from the 2 sites with unequal powers. It is a real case in SFN where the terminal receives signals from the 2 sites transmitters. We will assume that the relative power imbalance factor between the received signals from the two sites is equal to β. At the receiving side, we assume that a sub-optimal iterative receiver is used for non-orthogonal STBC schemes. The sub-optimal solution proposed here consists of an iterative receiver where the ST detector and the channel decoder exchange extrinsic information in an iterative way until the algorithm converges [2].

## 2.1 Single layer case: inter-cell ST coding

In the single layer case i.e. $M_T$=1, the second layer matrix $\mathbf{X}^{(2)}$ resumes to one element. The MIMO transmission is therefore achieved by the set of one antenna in each site. Due to the mobility of the terminal i.e. different assumed positions, the first layer ST scheme must be efficient face to unequal received powers. In this letter, we consider the orthogonal Alamouti code [3], the space multiplexing (SM) scheme [4] and the Golden code [5] with $M_R$= 2 Rx antennas. Without loss of generality, we assume that the transmission from a transmitting antenna *i* to a receiving antenna *j* is achieved for each sub-carrier *n* through a frequency non-selective Rayleigh fading channel.

Figure 2 gives the required $E_b/N_0$ to obtain a bit error rate (BER) equal to $10^{-4}$ for different values of β and a spectral efficiency η=4 [b/s/Hz]. As expected, this figure shows that the Golden code presents the best performance when the 2 Rx antennas receive the same power from the 2 sites (i.e. β=0 dB). When β decreases however, the Alamouti scheme is the most efficient since it presents only 3 dB loss in terms of required $E_b/N_0$ with respect to the case of equal received powers. Indeed, the transmission scenario becomes equivalent to a transmission scenario with one Tx antenna for very small values of β.

## 2.2 Double layer case

In the case of a double layer reception, the code construction is based on the single layer results. We restrict our study to $M_T=2$ Tx antennas by site and $M_R=2$ Rx antennas. We construct the first layer with the Alamouti scheme, since it is the most resistant for the case of unequal received powers. In a complementary way, we propose to construct the second layer with the Golden code since it offers the best results in the case of equal received powers. After combination of the 2 layers, (1) yields:

$$X = \frac{1}{\sqrt{5}} \begin{pmatrix} \alpha(s_1+\theta s_2) & \alpha(s_3+\theta s_4) & \alpha(s_5+\theta s_6) & \alpha(s_7+\theta s_8) \\ j\bar{\alpha}(s_3+\bar{\theta}s_4) & \bar{\alpha}(s_1+\bar{\theta}s_2) & j\bar{\alpha}(s_7+\bar{\theta}s_8) & \bar{\alpha}(s_5+\bar{\theta}s_6) \\ -\alpha^*(s_5^*+\theta^* s_6^*) & -\alpha^*(s_7^*+\theta^* s_8^*) & \alpha^*(s_1^*+\theta^* s_2^*) & \alpha^*(s_3^*+\theta^* s_4^*) \\ j\bar{\alpha}^*(s_7^*+\bar{\theta}^* s_8^*) & -\bar{\alpha}^*(s_5^*+\bar{\theta}^* s_6^*) & -j\bar{\alpha}^*(s_3^*+\bar{\theta}^* s_4^*) & \bar{\alpha}^*(s_1^*+\bar{\theta}^* s_2^*) \end{pmatrix} \quad (2)$$

where $\theta = \frac{1+\sqrt{5}}{2}, \bar{\theta}=1-\theta, \alpha=1+j(1-\theta), \bar{\alpha}=1+j(1-\bar{\theta})$, $\mu=j$ and $(.)^*$ stands for complex conjugate.

Figure 3 shows the results in terms of required $E_b/N_0$ to obtain a BER equal to $10^{-4}$ for different values of β and 3 STBC schemes i.e. our proposed 3D code scheme, the 1-Layer Alamouti and the Golden code schemes.

Figure 3 shows that our proposed scheme presents the best performance whatever the spectral efficiency and the factor β. Indeed, it is optimized for SFN systems owing to the robustness of the Alamouti scheme to unbalanced received powers and the full rank of the Golden code. For β=-12 dB, the proposed 3D code offers a gain equal to 1.8 dB (respectively 3 dB) with respect to the Alamouti scheme for η=4 [b/s/Hz] (resp. η=6 [b/s/Hz]). This gain is even greater when it is compared to the Golden code. Moreover, the maximum loss of our code due to unbalanced received powers is only equal to 3 dB in terms of $E_b/N_0$. These results confirm that the proposed 3D code is very robust whatever the spectral efficiency and the imbalance factor β. Eventually, we should note that the factor β could be related to the channel impulse response delay and to the power path loss. Then, it can be used to adjust synchronisation problems.

## 3 Conclusion

In this letter, a new 3D STSBC is presented. It is based on a double layer structure defined for inter-cell and intra-cell situations by adequately combining the Alamouti code and the Golden code performance. We showed that our proposed scheme is very efficient to cope with equal and unequal received powers in SFN scenarios.

## Acknowledgments

The authors would like to thank the European CELTIC project "B21C" for its support of this work.## References

[1] A. Mattson, "Single frequency networks in DTV", IEEE Trans. on Broadcasting, Vol. 51, Issue 4, pp.: 413-422, Dec. 2005.

[2] Y. Nasser, J.-F. Hélard, and M. Crussiere, "On the Influence of Carrier Frequency Offset and Sampling Frequency Offset in MIMO-OFDM Systems for Future Digital TV", in the proceedings of the IEEE International Symposium on Wireless and Pervasive computing, pp. 93-96, May 2008, Santorini, Greece.

[3] S.M. Alamouti, "A simple transmit diversity technique for wireless communications", IEEE J. on Selected Areas in Communications, vol. 16, no. 8, pp. 1451-1458, Oct. 1998.

[4] G. J. Foschini, "Layered space-time architecture for wireless communication in a fading environment when using multi-element antenna," Bell Labs Tech. J., vol. 1, no. 2, pp. 41–59, 1996.

[5] J.-C. Belfiore, G. Rekaya, and E. Viterbo, "The golden code: a 2 × 2 full-rate space-time code with non-vanishing determinants," IEEE Trans. in Information Theory, vol. 51, no. 4, pp. 1432–1436, Apr. 2005.7

**Author's affiliation**

Youssef Nasser, *member IEEE*, youssef.nasser@insa-rennes.fr, (Institute of Electronics and Telecommunications of Rennes, INSA Rennes, 20 Avenue de Buttes des Coesmes, 35043 Rennes cedex, France).

Jean-François Hélard, *Senior member IEEE*, jean-francois.helard@insa-rennes.fr, (Institute of Electronics and Telecommunications of Rennes, INSA Rennes, 20 Avenue de Buttes des Coesmes, 35043 Rennes cedex, France).

Matthieu Crussière, *member IEEE*, matthieu.crussiere@insa-rennes.fr, (Institute of Electronics and Telecommunications of Rennes, INSA Rennes, 20 Avenue de Buttes des Coesmes, 35043 Rennes cedex, France).

**Figure Caption**

Figure 1- SFN network with unequal received powers ($M_T$=2)

Figure 2- Required Eb/N0 to obtain a BER=$10^{-4}$, single layer case, DVB-T parameters**,** convolutional encoder (171,133)$_o$

$\eta$=4 [b/s/Hz] : Alamouti code: 64-QAM, R=1, channel encoding rate $R_c$=2/3.

Other schemes: 16-QAM, R=2, $R_c$ =1/2.

Figure 3- Required Eb/N0 to obtain a BER=$10^{-4}$, double layer case, DVB-T parameters,

η=4 [b/s/Hz]: Alamouti code: 64-QAM, R=1, $R_c$ =2/3.

            Other schemes: 16-QAM, R=2, $R_c$ =1/2.

η=6 [b/s/Hz]: Alamouti code: 256-QAM, R=1, $R_c$ =3/4.

            Other schemes: 64-QAM, R=2, $R_c$ =1/2.

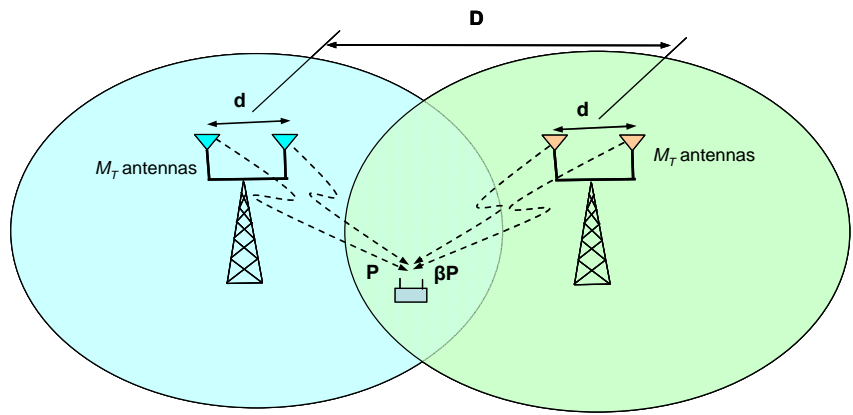

Figure 1 :

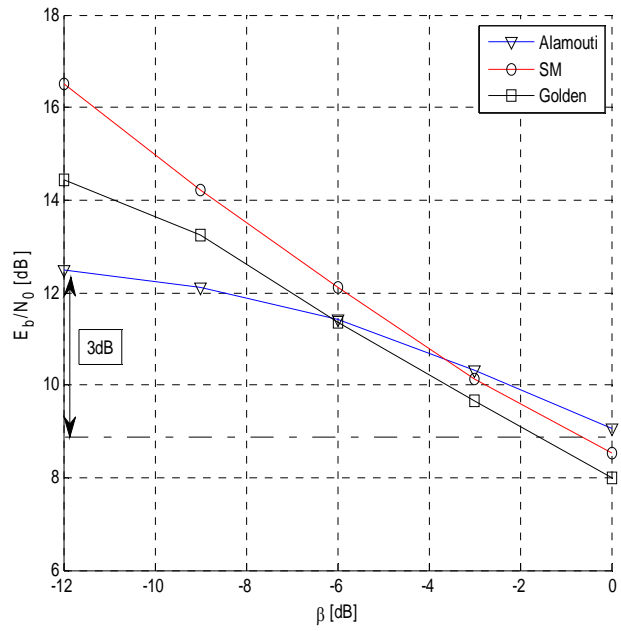

Figure 2 :

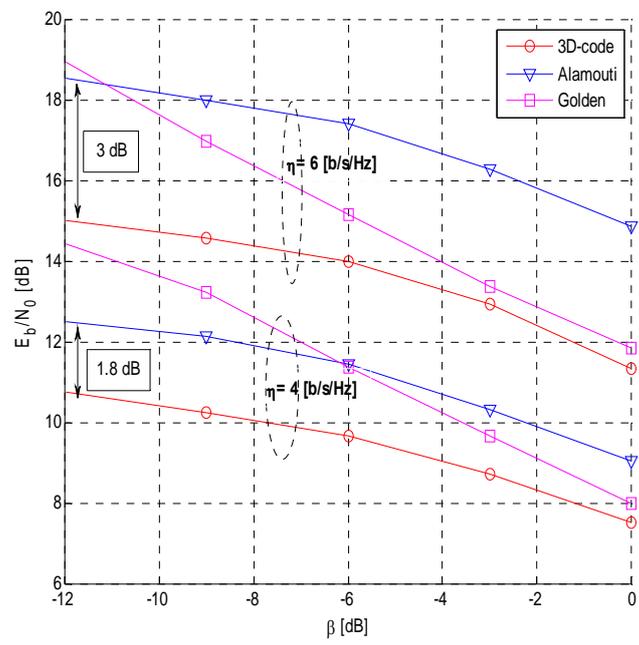

Figure 3 :

10